\documentclass[preprintnumbers,amsmath,12pt,amssymb,floatfix,superscriptaddress,nofootinbib]{revtex4}
\usepackage{graphicx}
\usepackage{epsfig}
\usepackage{bm}
\usepackage{amsfonts}
\def\be {\begin{equation}}
\def\ee {\end{equation}}
\def\ba {\begin{eqnarray}}
\def\ea {\end{eqnarray}}

\begin{document}
\title{Power-Law Entropy Corrected Ricci Dark Energy and Dynamics
of  Scalar Fields} 

 \author{\textbf{Antonio Pasqua}}\email{toto.pasqua@gmail.com}
  \affiliation{Department of Physics, University of Trieste, Trieste, Italy}

 \author{\textbf{Mubasher Jamil}}\email{mjamil@camp.nust.edu.pk}
\affiliation{Center for Advanced Mathematics and Physics (CAMP),
National University of Sciences and Technology (NUST), H-12,
Islamabad, Pakistan}\affiliation{Eurasian International Center
for Theoretical Physics, Eurasian National University, Astana
010008, Kazakhstan}

 \author{\textbf{Ratbay Myrzakulov}}\email{rmyrzakulov@gmail.com;
rmyrzakulov@csufresno.edu}\affiliation{Eurasian International Center
for Theoretical Physics, Eurasian National University, Astana
010008, Kazakhstan}

 \author{\textbf{Bushra Majeed}}\affiliation{Center for Advanced Mathematics and Physics (CAMP),
National University of Sciences and Technology (NUST), H-12,
Islamabad, Pakistan}

\begin{abstract}
\textbf{Abstract:} Motivated by the holographic principle, it has been suggested that the Dark Energy (DE) density can be inversely proportional to the area $A$ of
the event horizon of the universe. However, this kind of model would have a casuality problem. In this work, we study the power-law entropy corrected holographic
DE (PLECHDE) model in the non-flat Friedmann-Robertson-Walker universe, with the  future event
horizon replaced by the average radius of the Ricci scalar curvature. We derive the equation of state parameter $\omega_{\Lambda}$, the deceleration parameter $q$
and the evolution of energy density parameter $\Omega_D'$ in presence of interaction between DE and Dark Matter (DM). We consider the correspondence
between our Ricci-PLECHDE model and the Modified Chaplygin Gas (MCG) and the tachyon, K-essence, dilaton and quintessence scalar fields.
The potential and the dynamics of the scalar field models have been reconstructed according to the evolutionary behaviour of the interacting entropy-corrected
holographic DE model.
\end{abstract}

\maketitle
\newpage

\section{Introduction}
Cosmological observations like the Supernovae Ia (SNeIa), the Cosmic Microwave Background (CMB) radiation anisotropies, the Large Scale Structure (LSS) and X-ray
experiments support the evidence for an accelerated expansion of our universe \cite{1}. A missing energy component with negative pressure is considered by astrophysicists
and cosmologists as responsable of this accelerated expansion. This missing component is also known as Dark Energy (DE). Recent analysis of cosmological observations
indicates that the two-thirds of the total energy of the universe is been occupied by the DE whereas DM occupies almost the remaining part (the baryonic matter we
observe represents only a few percent of the total mass of the universe) \cite{1-1}. The contribution of the radiation is negligible.\\
The nature of DE is still unkwown and many candidates have been proposed in order to describe it  \cite{2}.
The simplest candidate for DE is a tiny positive cosmological constant, with a negative constant EoS parameter $\omega$, i.e. $\omega=-1$.
However cosmologists know that the cosmological
constant suffers from two well-known difficulties, the fine-tuning and the cosmic coincidence problems: the former asks why the vacuum energy density is so small
(of the order of $10^{-123}$ smaller than what we observe)
and the latter says why vacuum energy and DM are nearly equal today (which represents an incredible coincidence if internal connections between them does not exist) \cite{2-2}.\\
As possible alternative to cosmological constant, dynamical scalar field models have been
proposed some of which are quintessence \cite{3}, phantom \cite{4}, f-essence \cite{fff} and K-essence \cite{5}. \\
An important advance in the studies of black hole theory and string theory is the suggestion
of the so called holographic principle which was proposed by Fischler and Susskind in 1998 \cite{8}. According to the holographic principle, the number
of degrees of freedom of a physical system should be finite and should scale with its
bounding area rather than with its volume \cite{thooft} and it should be constrained
by an infrared cut-off \cite{12}.
The Holographic DE (HDE), based on the holographic principle, is one of the most studied models of DE \cite{11}. HDE
models have also been constrained and tested by various astronomical observation \cite{const} and by the anthropic principle \cite{antro}.\\
Applying the holographic principle to cosmology, the upper bound of the entropy contained
in the universe can be obtained. Following this line, Li \cite{10} suggested as constraint on the energy density of the universe $\rho_\Lambda\leq 3\gamma M^2_pL^{-2}$, where $\gamma$ is a numerical constant,
$L$ is the IR cut-off radius and $M_p=(8\pi G)^{-\frac{1}{2}}\simeq 10^{18} GeV$ is the reduced Planck mass. The equality sign holds when the holographic bound is
saturated. Since the definition and the derivation of the HDE density depends on the entropy-area relation $S\sim A\sim L^2$ of the black hole, where $A$ is the area
of the horizon, we can modify the definition of HDE taking into account power-law corrections to the entropy. These corrections appear in dealing with the
entanglement of quantum fields in and out the horizon \cite{13}. The form of the power-law entropy is given by \cite{14}:
\begin{eqnarray}
S=\frac{A}{4G}\left(1-K_\alpha A^{1-\frac{\alpha}{2}}\right), \label{1}
\end{eqnarray}
where $\alpha$ is a dimensionless constant which value is still unknown and:
\begin{eqnarray}
K_\alpha=\frac{\alpha(4\pi)^{\frac{\alpha}{2}-1}}{(4-\alpha)r^{2-\alpha}_c}. \label{2}
\end{eqnarray}
The quantity $r_c$ represents the crossover scale.\\
The second term in Eq. (\ref{1}) is regarded as a power-law correction to the area law. Inspired by the power-law corrected entropy relation given in Eq. (\ref{1}), the energy density
$\rho_{\Lambda}$ of the PLECHDE is obtained as \cite{shejam}:
\begin{eqnarray}
\rho_{\Lambda}=3\gamma M^2_pL^{-2}-\beta M^2_pL^{-\alpha}, \label{3}
\end{eqnarray}
where $\gamma$ and $\beta$ are two constants of the order of the unity. In the limiting case of $\beta=0$, we get the ordinary HDE density.  \\
In this paper, we propose the R-PLECHDE model which is obtained by using as IR cut-off radius the average radius of Ricci scalar curvature, i.e. $L=R^{-\frac{1}{2}}$. For a
non-flat universe, the Ricci scalar $R$ is given by:
\begin{equation}
 R=6\left(\dot{H}+2H^2+\frac{k}{a\left(t\right)^2}\right),\label{4}
\end{equation}
where $H=\frac{\dot{a}}{a}$ is the Hubble parameter, $\dot{H}$ is the derivative of the Hubble parameter with respect to the cosmic time $t$, $a\left(t\right)$ is a
dimensionless scale factor (which is function of the cosmic time) and $k$ is the curvature parameter which can assume the values $-1,\, 0,\, +1$ which yield, respectively, a closed, a flat or an open FRW
universe. The curvature paramater $k$ has dimension of $length^{-2}$ and it describes the spatial geometry of space-time.\\
The average radius of the Ricci scalar curvature was proposed for the first time as infrared cut-off by Gao et al. \cite{gao}.
It was found that this kind of model works well when observational data are fitted and it can also be helpful to understand the coincidence problem.
Moreover, the presence of the event horizon is not presumed in this model, so it is possible to avoid the casuality problem.  \\
Thanks to the work of Cai, Hu and Zhang \cite{CAI}, which studied the casual entropy bound in the holographic framework, the Ricci model gets an appropriate reason for which it could be motivated, providing an appropriate physical motivation for the holographic Ricci DE (RDE). \\
RDE has been widely studied in literature in various ways: the statefinder diagnostic of RDE \cite{feng1},reconstruction of $f\left(R \right)$ \cite{feng2},
quintom \cite{feng4},contributions of viscosity to RDE \cite{feng3}, and related observational constraints \cite{xu}. \\
Replacing $L$ with $R^{-1/2}$ in Eq. (\ref{3}), we get the energy density of R-PLECHDE as:
\begin{equation}
\rho_\Lambda=3\gamma M^2_pR-\beta M^2_pR^{\frac{\alpha}{2}}.\label{5}
\end{equation}
which is an extension of Ricci DE model proposed by Gao.\\
This paper is organized as follows. In Section 2, we describe the physical context we are working in and the R-PLECHDE model is described; moreover, we derive the EoS parameter $\omega_{\Lambda}$, the deceleration parameter $q$ and the
evolution of the energy density paramater $\Omega'_{\Lambda}$. In Section 3, we establish a correspondence between R-PLECHDE model and the Modified Chaplygin Gas (MCG) and the tachyon, K-essence, dilaton and quintessence scalar fields. In Section 4, Conclusions are discussed. 

\section {The Model of R-PLECHDE}
Since cosmological observations show that our universe is not perfectly flat but it has a small positive curvature which implies a closed universe,
we consider in this paper a non-flat universe, then we will work in the FRW universe background. The tendency for a closed universe is obtained in different independet
cosmological experiments \cite{sperge2}. The line element for non-flat FRW universe is given by:
\begin{eqnarray}
    ds^2=-dt^2+a^2\left(t\right)\left(\frac{dr^2}{1-kr^2} +r^2 \left(d\theta ^2 + \sin^2 \theta d\varphi ^2\right) \right),\label{6}
\end{eqnarray}
where $t$ is the cosmic time, $r$ is referred to the radial component and $\left( \theta, \phi\right)$ are the angular coordinates. \\
The Friedmann equation for non-flat FRW universe dominated by DE and DM takes the form:
\begin{eqnarray}
    H^2+\frac{k}{a^2}=\frac{1}{3M^2_p}\left( \rho _{\Lambda} + \rho _{m}\right),\label{7}
\end{eqnarray}
where $\rho _{\Lambda}$ and $\rho _{m}$ are the energy densities of DE and DM, respectively.\\
We also define the fractional energy densities for DM, curvature and DE, respectively, as:
\begin{eqnarray}
    \Omega _m &=& \frac{\rho_m}{\rho _{cr}} = \frac{\rho _m}{3M_p^2H^2},\label{8}\\
    \Omega _k &=& \frac{\rho_k}{\rho _{cr}} = \frac{k}{H^2a^2},\label{9}\\
    \Omega _{\Lambda} &=& \frac{\rho_{\Lambda}}{\rho _{cr}} = \frac{\rho _{\Lambda}}{3M_p^2H^2},\label{10}
\end{eqnarray}
where $\rho_{cr} = 3M^2_pH^2$ represents the critical energy density. $\Omega _k$ represents the contribution to the total density from
the spatial curvature.  Recent observations reveal that $\Omega _k \cong 0.02$ \cite{sperge}, which support a closed universe with a small positive curvature.\\
Using the Friedmann equation given in Eq. (\ref{7}),  Eqs. (\ref{8}), (\ref{9}) and (\ref{10}) yield to:
\begin{eqnarray}
\Omega_{m} + \Omega_{\Lambda} = 1 + \Omega_k.\label{11}
\end{eqnarray}
In order to preserve the Bianchi identity or the local energy-momentum conservation law, i.e. $\nabla_{\mu}T^{\mu \nu}=0$, the total energy density $\rho_{tot}= \rho_D + \rho_m$
must satisfy the following relation:
\begin{eqnarray}
    \dot{\rho}_{tot}+3H\left( 1+\omega \right)\rho_{tot}=0,\label{12}
\end{eqnarray}
where $\omega \equiv p_{tot}/\rho_{tot}$ is the total EoS parameter. Since we are considering the interaction between DE and DM, the two energy densities
$\rho_\Lambda$ and $\rho_m$ are preserved separately and the equations of conservation assume the forms:
\begin{eqnarray}
    \dot{\rho}_{\Lambda}&+&3H\rho_{\Lambda}\left(1+\omega_{\Lambda}\right)=-Q, \label{13} \\
\dot{\rho}_m&+&3H\rho_m=Q, \label{14}
\end{eqnarray}
where $Q$ represents an interaction term which can be an arbitrary function of cosmological parameters, like the Hubble parameter $H$ and energy densities $\rho_m$
and $\rho_{\Lambda}$, i.e.
$Q(H\rho_m,H\rho_{\Lambda})$. The simplest and most used expression for $Q$ is given by:
\begin{eqnarray}
    Q = 3b^2H(\rho _m + \rho _{\Lambda}),\label{15}
\end{eqnarray}
where $b^2$ represents a coupling parameter between DE and DM \cite{17}. If $b^2>0$ we have transition from DE to DM, instead $b^2<0$ implies transition from DM to DE.
The case corresponding to $b^2=0$ represents the non-interacting FRW model, instead $b^2=1$ yields a complete transfer from DE to DM.
Recently, it was reported that this interaction is observed in the Abell cluster A586 showing a transition of DE into DM and vice versa \cite{abell1}.
However the strength of this interaction is not clearly identified \cite{abell2}.  \\
Observations of CMB and galactic clusters show that the coupling parameter is $b^2<0.025$, i.e. a small positive constant of the order of the unity \cite{19}. A
negative coupling constant results in the violation of thermodynamical laws so its avoided. We must also note that the ideal interaction term must be motivated from the
quantum gravity theory, otherwise we rely on dimensional basis for choosing an interaction term $Q$.
However, more general phenomenological terms can be used since the nature of DE and DM is still not well-understood. For this reason, different lagrangians have been
proposed in order to describe $Q$.\\
We now want to derive the EoS parameter $\omega_\Lambda$ for the R-PLECHDE model.
Using the Friedmann equation given in Eq. (\ref{7}), the Ricci scalar $R$ given in Eq. (\ref{4}) can be rewritten in the following form:
\begin{eqnarray}
    R=6\left(  \dot{H} + H^2 + \frac{\rho_m+\rho_{\Lambda}}{3M_p^2} \right).\label{18}
\end{eqnarray}
From the Friedmann equation given in Eq. (\ref{7}), we can also derive that:
\begin{eqnarray}
    \dot{H}=\frac{k}{a^2}-\frac{1}{2M_p^2}\left[  \rho_m+\rho_{\Lambda}\left( 1 + \omega_{\Lambda} \right) \right].\label{19}
\end{eqnarray}
Adding Eqs. (\ref{7}) and (\ref{19}), we obtain:
\begin{eqnarray}
    \dot{H}+H^2=\frac{\rho_m+\rho_{\Lambda}}{3M_p^2}-\frac{1}{2M_p^2}\left[  \rho_m+\rho_{\Lambda}\left( 1 + \omega_{\Lambda} \right) \right].\label{20}
\end{eqnarray}
So, the Ricci scalar $R$ given in Eq. (\ref{18}) can be rewritten as:
\begin{eqnarray}
 R=\frac{\rho_m+\rho_{\Lambda}}{M_p^2}-\frac{3\rho_{\Lambda}\omega_{\Lambda}}{M_p^2}.\label{!21!}
\end{eqnarray}
The EoS parameter $\omega_{\Lambda}$ can be now easily obtained from Eq. (\ref{!21!}) as follow:
\begin{eqnarray}
 \omega_{\Lambda}=-\frac{RM_p^2}{3\rho_{\Lambda}}+\frac{\Omega_{\Lambda}+\Omega_{m}}{3\Omega_{\Lambda}}, \label{22}
\end{eqnarray}
where we used the relation $\frac{\rho_{\Lambda}+\rho_{m}}{3\rho_{\Lambda}} = \frac{\Omega_{\Lambda}+\Omega_{m}}{3\Omega_{\Lambda}}$.\\
Substituting in Eq. (\ref{22}) the expression of the energy density $\rho_{\Lambda}$ of the R-PLECHDE given in Eq. (\ref{5}) and using Eq. (\ref{11}), we get:
\begin{eqnarray}
    \omega_{\Lambda}=-\frac{1}{3(3\gamma-\beta R^{\frac{\alpha}{2}-1})} + \frac{\left(1+ \Omega _k  \right)}{3\Omega _{\Lambda}},\label{23}
\end{eqnarray}
which represents the EoS parameter of the R-PLECHDE model.\\
We now want to derive the expression for the evolution of energy density parameter $\Omega_{\Lambda}$.\\
From Eq. (\ref{13}), we can obtain the following expression for the EoS parameter  $\omega_{\Lambda}$:
\begin{eqnarray}
    \omega_{\Lambda}= -1-\frac{\dot{\rho}_{\Lambda}}{3H\rho_{\Lambda}}-\frac{Q}{3H\rho_{\Lambda}}.\label{!23!}
\end{eqnarray}
Using the expression of $Q$ given in Eq. (\ref{15}), the derivative of the DE energy density $\rho_{\Lambda}$ with respect to the cosmic time can be written as:
\begin{eqnarray}
    \dot{\rho}_{\Lambda}=3H\left[ -\rho_{\Lambda}-\left(\rho_m + \rho_{\Lambda}\right)\left(b^2+\frac{1}{3} \right) + \frac{RM_p^2}{3} \right].\label{24-1}
\end{eqnarray}
Dividing Eq. (\ref{24-1}) by the critical density $\rho_c=3H^2M_p^2$, we obtain:
\begin{eqnarray}
    \frac{\dot{\rho}_{\Lambda}}{\rho_c}=3H\left[ -\Omega_{\Lambda}-\left(1 + \Omega_k\right)\left(b^2+\frac{1}{3} \right) + \frac{R}{9H^2} \right]=\dot{\Omega_{\Lambda}}+2\Omega_{\Lambda}\frac{\dot{H}}{H}.  \label{25-}
\end{eqnarray}
From Eq. (\ref{4}), we can derive:
\begin{eqnarray}
    \frac{R}{9H^2}=\frac{2}{3}\left( \frac{\dot{H}}{H^2} +2 + \Omega_k \right).\label{26-}
\end{eqnarray}
Substituting Eq. (\ref{26-}) in Eq. (\ref{25-}), it is possible to obtain the derivative of $\Omega_{\Lambda}$ with respect to the cosmic time $t$:
\begin{eqnarray}
\dot{\Omega}_{\Lambda}=2\frac{\dot{H}}{H}\left(1-\Omega_{\Lambda}  \right)+3H\left[-\Omega_{\Lambda}-  \left(1 + \Omega_k\right)\left(b^2-\frac{1}{3} \right) + \frac{2}{3} \right].\label{27-}
\end{eqnarray}
Since   $\Omega_{\Lambda}'=\frac{d\Omega_{\Lambda}}{dx}= \frac{1}{H}\dot{\Omega}_{\Lambda}$ (where $x=\ln a$), we derive:
\begin{eqnarray}
H   \Omega_{\Lambda}'=2H'\left(1-\Omega_{\Lambda}  \right)+3H\left[-\Omega_{\Lambda}-  \left(1 + \Omega_k\right)\left(b^2-\frac{1}{3} \right) + \frac{2}{3} \right],\label{28-}
\end{eqnarray}
which yields to:
\begin{eqnarray}
    \Omega_{\Lambda}'=\frac{2}{H}\left(1-\Omega_{\Lambda}  \right)+3\left[-\Omega_{\Lambda}-  \left(1 + \Omega_k\right)\left(b^2-\frac{1}{3} \right) + \frac{2}{3} \right].\label{29-}
\end{eqnarray}
In Eq. (\ref{29-}) we used the fact that:
\begin{eqnarray}
    H'=\frac{a'}{a}=1.\label{30}
\end{eqnarray}
For completeness, we now derive the expression of the deceleration parameter $q$, which is defined as:
\begin{eqnarray}
    q=-\frac{\ddot{a}a}{\dot{a}^2}=  -\frac{\ddot{a}}{aH^2}  =-1-\frac{\dot{H}}{H^2}. \label{24}
\end{eqnarray}
The deceleration paramater, combined with the Hubble parameter $H$ and the dimensionless density parameters, form a set of very useful parameters for the description
of the astrophysical observations. Taking the time derivative of the Friedmann given in Eq. (\ref{7}) and using Eqs. (\ref{11}) and (\ref{14}), it is possible to write
the deceleration parameter $q$ as:
\begin{eqnarray}
    q=\frac{1}{2}\left[1 + \Omega_k + 3\Omega_{\Lambda} \omega_{\Lambda}  \right].\label{25}
\end{eqnarray}
Substituting in Eq. (\ref{25}) the expression of the EoS parameter $\omega_{\Lambda}$ of the R-PLECHDE given in Eq. (\ref{23}), we obtain that:
\begin{eqnarray}
    q=1-\frac{\Omega_{\Lambda}}{2\left(3\gamma -\beta R^{\frac{\alpha} {2}-1}\right)}+ \Omega_k .\label{26}
\end{eqnarray}
We can now derive the important quantities of the R-PLECHDE model in the limiting case, for a flat dark dominated universe, i.e. when $\beta=0$, $\Omega_{\Lambda}=1$
and $\Omega_k$=0.\\
The energy density $\rho_{\Lambda}$ given in Eq. (\ref{5}) reduces to:
\begin{eqnarray}
\rho_{\Lambda}=3\gamma M_{p}^2 R.\label{27}
\end{eqnarray}
From the Friedmann equation given in Eq. (\ref{7}), we can derive the following expressions for the Hubble parameter $H$ and the Ricci scalar curvature $R$:
\begin{eqnarray}
H=\frac{6\gamma}{12\gamma-1}\left(\frac{1}{t}\right),\label{28}
\end{eqnarray}
\begin{eqnarray}
R=\frac{36\gamma}{(12\gamma-1)^2}\left(\frac{1}{t^2}\right).\label{29}
\end{eqnarray}
Finally, the EoS parameter $\omega_{\Lambda}$ and deceleration parameter $q$ reduce, respectively, to:
\begin{eqnarray}
\omega_{\Lambda}=\frac{1}{3}-\frac{1}{9\gamma},\label{30}
\end{eqnarray}
\begin{eqnarray}
q=1-\frac{1}{6\gamma}.\label{31}
\end{eqnarray}
From Eq. (\ref{30}), we see that in this limit, the EoS parameter of DE become a constant value and for $\gamma<1/12$ we have $\omega_{\Lambda}<-1$, where the phantom
divide can be crossed. Since the Ricci scalar $R$ given in Eq. (\ref{29}) diverges at $\gamma=1/12$, this value can not be taken into account. From Eq. (\ref{31}), we obtain that
the acceleration starts at $\gamma\leq 1/6$, where the quintessence regime is started ($\omega_{\Lambda} \leq -1/3$).\\
The result obtained is very similar to the power-law expansion of scale factor found by Granda and Oliveros in 2008 \cite{granda2008},
in which $a(t)=t^{6\gamma /(12\gamma-1)}$.

\section{CORRESPONDENCE BETWEEN R-PLECHDE AND SCALAR FIELDS}
In this Section, we establish a correspondence between the interacting Ricci power-law corrected model and the tachyon, K-essence, dilaton and quintessence scalar field
models and the Modified Chaplygin Gas (MCG). The importance of this correspondence is that the scalar field models are an effective description of an underlying
theory of DE. Therefore, it is worthwhile to reconstruct the potential and the dynamics of scalar fields according the evolutionary form of Ricci scalar model. For
this purpose, first we compare the energy density of Ricci scale model given in Eq. (\ref{5}) with the energy density of corresponding scalar field model. Then, we
equate the EoS parameters of scalar field models with the EoS parameter of Ricci scalar model given in Eq. (\ref{23}).

\subsection{INTERACTING TACHYON MODEL}
Recently, a huge interest has been devoted to the study of the inflationary model with the help of the tachyon field, since it is believed the tachyon can be assumed
as a possible source of DE \cite{20}. The tachyon is an unstable field which can be used in string theory through its role in the Dirac-Born-Infeld (DBI) action to
describe the D-brane action \cite{21}. Tachyon might be responsible for cosmological inflation in the early evolutionary stage of the universe, due to tachyon
condensation near the top of the effective scalar potential. A rolling tachyon has an interesting EoS whose parameter smoothly interpolates in the range [$-1$,0].
This discovery motivated to take DE as a dynamical quantity, i.e. a variable cosmological constant and model inflation using tachyons.\\
The effective Lagrangian for the tachyon field is given by:
\begin{eqnarray}
L=-V(\phi)\sqrt{1-g^{\mu \nu}\partial _{\mu}\phi \partial_{\nu}\phi},\label{32}
\end{eqnarray}
where $V(\phi)$ represents the potential of tachyon and $g^{\mu \nu}$ is the metric tensor. The energy density $\rho_{\phi}$ and pressure  $p_{\phi}$ for the tachyon field are given, respectively, by:
\begin{eqnarray}
    \rho_{\phi}&=&\frac{V(\phi)}{\sqrt{1-\dot{\phi}^2}}, \label{33} \\
p_{\phi}&=& -V(\phi)\sqrt{1-\dot{\phi}^2}. \label{34}
\end{eqnarray}
The EoS parameter $\omega_{\phi}$ of tachyon scalar field can be obtained from the following expression:
\begin{eqnarray}
\omega_{\phi}=\frac{p_{\phi}}{\rho_{\phi}}=\dot{\phi}^2-1. \label{35}
\end{eqnarray}
In order to have a real energy density for tachyon field, it is required that $-1 < \dot{\phi} < 1$. Consequently, from Eq. (\ref{35}), the EoS parameter of tachyon
is constrained in the range $-1 < \omega_{\phi} < 0$. Hence, the tachyon field can interpret the accelerated expansion of the universe, but it can not enter the
phantom regime which has $\omega_{\Lambda}<-1$.\\
Comparing Eqs. (\ref{5}) and (\ref{33}), we obtain the following expression for the potential $V(\phi)$ of the tachyon field:
\begin{eqnarray}
    V(\phi)=\rho_{\Lambda} \sqrt{1-\dot{\phi}^2}. \label{36}
\end{eqnarray}
Instead, equating Eqs. (\ref{23}) and (\ref{35}), we derive the expression of the kinetic energy term $\dot{\phi}^2$ as follow:
\begin{eqnarray}
\dot{\phi}^2= 1 + \omega_{\Lambda}=1-\frac{1}{3\left(3\gamma -\beta
R^{\frac{\alpha}{2}-1}\right)} + \frac{\left(1+ \Omega _k \right)}{3\Omega
_{\Lambda}}.\label{37}
\end{eqnarray}
Moreover, inserting Eq. (\ref{37}) into Eq. (\ref{36}), it is possible to write the potential $V\left(\phi \right)$ of the tachyon as follow:
\begin{eqnarray}
    V\left( \phi  \right) &= &\rho _{\Lambda} \sqrt{\frac{1}{3\left(3\gamma -\beta R^{\frac{\alpha}{2}-1}\right)} -
    \frac{\left(1+ \Omega _k  \right)}{3\Omega _{\Lambda}}}.\label{38}
\end{eqnarray}
Since $\dot{\phi}=\phi'H$, from Eq. (\ref{37}) it follows that:
\begin{eqnarray}
 \phi'= \frac{1}{H}  \sqrt{1-\frac{1}{3\left(3\gamma -\beta R^{\frac{\alpha}{2}-1}\right)}+ \frac{\left(1+ \Omega _k  \right)}{3\Omega _{\Lambda}}}.\label{39}
\end{eqnarray}
It is now possible to derive the evolutionary form of the tachyon scalar field integrating Eq. (\ref{39}) with respect to the scale factor $a$:
\begin{eqnarray}
    \phi\left(a\right) - \phi\left(a_0\right)=\int_{a_0}^a \frac{da}{aH}\sqrt{1-\frac{1}{3\left(3\gamma -\beta R^{\frac{\alpha}{2}-1}\right)} + \frac{\left(1+ \Omega _k  \right)}{3\Omega _{\Lambda}}},\label{40}
\end{eqnarray}
where $a_0$ is the present value of the scale factor.\\
In the limiting case for flat dark dominated universe, i.e. when $\beta=0$, $\Omega_{\Lambda}=1$ and $\Omega_k$=0, the scalar field and potential of the tachyon become,
respectively:
\begin{eqnarray}
\phi(t)=\left(\sqrt{\frac{12\gamma -1}{9\gamma}}\right)t,\label{41}
\end{eqnarray}
\begin{eqnarray}
V(\phi)=\frac{4M_p^2\sqrt{\gamma\left(1-3\gamma\right)}}{(12\gamma-1)}\left(\frac{1}{\phi^2}\right).\label{42}
\end{eqnarray}
In this correspondence, the scalar field exist when $\gamma>1/12$, which shows that the phantom divide can not be achieved.

\subsection{INTERACTING K-essence MODEL}
A model in which the kinetic term of the scalar field appears in the Lagrangian in a non-canonical way is known as K-essence model. The idea of the K-essence scalar
field was motivated from the Born-Infeld action of string theory and it is used to explain the late time acceleration of the universe \cite{22}. The general scalar
field action $S_K$ for the K-essence field as a function of $\phi$ and $\chi=\dot{\phi}/2$ is given by \cite{23}:
\begin{eqnarray}
S_K=\int d^4x \sqrt{-g}\,p\left(\phi, \chi \right),\label{43}
\end{eqnarray}
The Lagrangian density $p\left(\phi, \chi \right)$ corresponds to a pressure density. According to Eq. (\ref{43}), the pressure
$p\left(\phi, \chi \right)$ and the energy density $\rho_{\Lambda}\left(\phi, \chi \right)$ of the K-essence can be written, respectively, as:
\begin{eqnarray}
    p\left(\phi, \chi \right)&=&f\left(\phi\right)\left( -\chi+\chi ^2   \right),\label{44} \\
        \rho\left(\phi, \chi \right)&=&f\left(\phi\right)\left(-\chi+3\chi ^2\right).\label{45}
\end{eqnarray}
The EoS parameter $\omega_K$ of K-essence scalar field is given, then, by:
\begin{eqnarray}
    \omega_K= \frac{p\left(\phi, \chi \right)}{\rho\left(\phi, \chi \right)}=\frac{\chi-1}{3\chi -1}.\label{46}
\end{eqnarray}
From Eq. (\ref{46}), we can see the phantom behaviour of K-essence scalar field ($\omega_K < -1$) is obtained when the parameter $\chi$ lies in the range
$1/3 < \chi < 1/2$.\\
In order to consider the K-essence field as a description of the interacting R-PLECHDE model, we establish the correspondence between the K-essence EoS parameter
$\omega_K$ and the R-PLECHDE EoS parameter $\omega_{\Lambda}$ given in Eq. (\ref{23}).\\
The expression for $\chi$ can be found equating  Eqs. (\ref{23}) and (\ref{46}), which yields to:
\begin{eqnarray}
    \chi = \frac{\omega_{\Lambda}-1}{3\omega_{\Lambda}-1}=\frac{-1-\frac{1}{3\left(3\gamma -\beta R^{\frac{\alpha}{2}-1}\right)}+ \frac{\left(1+ \Omega _k  \right)}
{3\Omega _{\Lambda}}}{-1- \frac{1}{\left(3\gamma -\beta R^{\frac{\alpha}{2}-1}\right)} + \frac{\left(1+ \Omega _k  \right)}{\Omega _{\Lambda}}}.\label{47}
\end{eqnarray}
Moreover, equating Eqs. (\ref{5}) and (\ref{45}), we derive:
\begin{eqnarray}
    f\left(\phi \right)=\frac{\rho_{\Lambda}}{\chi(3\chi-1)}.\label{48}
\end{eqnarray}
Using $\dot{\phi}^2=2\chi$ and the relation $\dot{\phi}=\phi'H$, we derive from Eq. (\ref{47}) that:
\begin{eqnarray}
 \phi'= \frac{\sqrt{2}}{H}\sqrt{\frac{-1-\frac{1}{3\left(3\gamma -\beta R^{\frac{\alpha}{2}-1}\right)}+ \frac{\left(1+ \Omega _k  \right)}{3\Omega _{\Lambda}}}{-1- \frac{1}{\left(3\gamma -\beta R^{\frac{\alpha}{2}-1}\right)} + \frac{\left(1+ \Omega _k  \right)}{\Omega _{\Lambda}}}}.\label{49}
\end{eqnarray}
We can now find the evolutionary form of the K-essence scalar field integrating Eq. (\ref{49}) with respect to the scale factor $a$:
\begin{eqnarray}
    \phi\left(a\right) -    \phi\left(a_0\right) = \sqrt{2} \int_{a_0}^a \frac{da}{aH}\sqrt{\frac{-1-\frac{1}{3\left(3\gamma -\beta R^{\frac{\alpha}{2}-1}\right)}+
 \frac{\left(1+ \Omega _k  \right)}{3\Omega _{\Lambda}}}{-1- \frac{1}{\left(3\gamma -\beta R^{\frac{\alpha}{2}-1}\right)} + \frac{\left(1+ \Omega _k  \right)}{\Omega
_{\Lambda}}}}, \label{50}
\end{eqnarray}
where $a_0$ is the present value of the scale factor. \\
In the limiting case for flat dark dominated universe, i.e. when $\beta=0$, $\Omega_{\Lambda}=1$ and $\Omega_k$=0, the scalar field and potential of K-essence field
reduce, respectively, to:
\begin{eqnarray}
     \phi(t)=\left(\sqrt{\frac{12\gamma+2}{3}}\right)t,\label{51}
\end{eqnarray}
\begin{eqnarray}
     f(\phi)=\frac{36\gamma M_p^2}{(12\gamma-1)^2}\left(\frac{1}{\phi^2}\right), \label{52}
\end{eqnarray}
which are a result of power-law expansion.\\
We see that the universe may behave in all accelerated regimes (i.e., phantom and quintessence), since all the values of $\gamma$ are possible. We can also note that
the results of this subsection can be extended for the g-essence as well as for f-essence \cite{myrzakulov}. This task is left for future investigations.

\subsection{INTERACTING DILATON MODEL}
A dilaton scalar field, originated from the lower-energy limit of string theory \cite{24}, can also be assumed as a source of DE.\\
The process of compactification of the string theory from higher to four dimensions introduces the scalar dilaton field which is coupled to curvature invariants.
The coefficient of the kinematic term of the dilaton can be negative in the Einstein frame, which means that the dilaton behaves as a phantom-like scalar field.
The pressure (Lagrangian) density and the energy density of the dilaton DE model are given, respectively, by \cite{25}:
\begin{eqnarray}
p_D&=&-\chi +ce^{\lambda \phi}\chi^2, \label{53}  \\
\rho_D&=&-\chi +3ce^{\lambda \phi}\chi^2.\label{54}
\end{eqnarray}
$c$ and $\lambda$ are two positive constants and $2\chi=\dot{\phi}^2$. The EoS parameter $\omega_D$ for the dilaton scalar field can be obtained from:
\begin{eqnarray}
    \omega _D= \frac{p_D}{\rho_D}=\frac{-1 +ce^{\lambda \phi}\chi}{-1 +3ce^{\lambda \phi}\chi}.\label{55}
\end{eqnarray}
In order to consider the dilaton field as a description of the interacting R-PLECHDE density, we now establish a correspondence between the dilaton EoS parameter
$\omega_D$ and the EoS parameter $\omega_{\Lambda}$ of the R-PLECHDE model given in Eq. (\ref{23}). By equating Eqs. (\ref{23}) and (\ref{55}), we find:
\begin{eqnarray}
    ce^{\lambda \phi}\chi = \frac{\omega_{\Lambda}-1}{3\omega_{\Lambda}-1}=\frac{-1-\frac{1}{3\left(3\gamma -\beta R^{\frac{\alpha}{2}-1}\right)}+ \frac{\left(1+ \Omega _k  \right)}{3\Omega _{\Lambda}}}{-1- \frac{1}{\left(3\gamma -\beta R^{\frac{\alpha}{2}-1}\right)} + \frac{\left(1+ \Omega _k  \right)}{\Omega_{\Lambda}}}.\label{56}
\end{eqnarray}
Since $\dot{\phi}^2=2\chi$, Eq. (\ref{56}) can be rewritten as:
\begin{eqnarray}
    e^{\lambda \phi/2} \dot{\phi}= \sqrt{\frac{2}{c}}\times \sqrt{\frac{-1-\frac{1}{3\left(3\gamma -\beta R^{\frac{\alpha}{2}-1}\right)}+ \frac{\left(1+ \Omega _k  \right)}{3\Omega _{\Lambda}}}{-1- \frac{1}{\left(3\gamma -\beta R^{\frac{\alpha}{2}-1}\right)} + \frac{\left(1+ \Omega _k  \right)}{\Omega_{\Lambda}}}}.\label{57}
\end{eqnarray}
Integrating Eq. (\ref{57}) with respect to the scale factor $a$, we obtain:
\begin{eqnarray}
    e^{\lambda \phi\left(a\right)/2} &=& e^{\lambda \phi\left(a_0\right)/2}+\frac{\lambda}{\sqrt{2c}}\int_{a_0}^a\frac{da}{aH}\sqrt{\frac{-1-\frac{1}{3\left(3\gamma
-\beta R^{\frac{\alpha}{2}-1}\right)}+ \frac{\left(1+ \Omega _k  \right)}{3\Omega _{\Lambda}}}{-1- \frac{1}{\left(3\gamma -\beta R^{\frac{\alpha}{2}-1}\right)} +
 \frac{\left(1+ \Omega _k  \right)}{\Omega_{\Lambda}}}},  \label{58}
\end{eqnarray}
where $a_0$ is the present value of the scale factor.\\
The evolutionary form of the dilaton scalar field is given by:
\begin{eqnarray}
    \phi\left(a\right)=  \frac{2}{\lambda}\log \left[e^{\lambda \phi \left(a_0\right)/2}+\frac{\lambda}{\sqrt{2c}}  \int_{a_0}^a \frac{da}{aH}\sqrt{ \frac{-1-\frac{1}{3\left(3\gamma -\beta R^{\frac{\alpha}{2}-1}\right)}+ \frac{\left(1+ \Omega _k  \right)}{3\Omega _{\Lambda}}}{-1- \frac{1}{\left(3\gamma -\beta R^{\frac{\alpha}{2}-1}\right)} + \frac{\left(1+ \Omega _k  \right)}{\Omega
    _{\Lambda}}}} \right].   \label{59}
\end{eqnarray}
In the limiting case for flat dark dominated universe, i.e. when $\beta=0$, $\Omega_{\Lambda}=1$ and $\Omega_k$=0, the scalar field of dilaton field reduces to:
\begin{eqnarray}
\phi(t)=\frac{2}{\lambda}\ln{\left[\lambda t\sqrt{\frac{1+6\gamma}{6c}}\right]}. \label{60}
\end{eqnarray}
We see from Eq. (\ref{60}) that $\gamma$ can assume all the possible values. Therefore, by this correspondence, the universe may behave both in phantom and
quintessence regime.

\subsection{QUINTESSENCE}
Quintessence is described by an ordinary time-dependent and homogeneous scalar field $\phi$ which is minimally coupled to gravity, but with a particular potential
$V\left(\phi\right)$ that leads to the accelerating universe. The action for quintessence is given by \cite{cope}:
\begin{eqnarray}
    S=\int d^4x \sqrt{-g}\,\left[-\frac{1}{2}g^{\mu \nu} \partial _{\mu} \phi   \partial _{\nu} \phi - V\left( \phi \right)  \right].\label{61}
\end{eqnarray}
The energy momentum tensor $T_{\mu \nu}$ of the quintessence field is derived by varying the action $S$ given in Eq. (\ref{61}) with respect to the metric tensor
$g^{\mu \nu}$:
\begin{eqnarray}
T_{\mu \nu}=\frac{2}{\sqrt{-g}} \frac{\delta S}{\delta g^{\mu \nu}},\label{62}
\end{eqnarray}
which yields, using Eq. (\ref{61}), to:
\begin{eqnarray}
    T_{\mu \nu}=\partial _{\mu} \phi   \partial _{\nu} \phi - g_{\mu \nu}\left[\frac{1}{2}g^{\alpha \beta} \partial _{\alpha} \phi   \partial _{\beta} \phi + V\left( \phi \right)  \right].\label{63}
\end{eqnarray}
Considering a FRW background, the energy density $\rho_Q$ and pressure $p_Q$ of the quintessence scalar field $\phi$ are given, respectively, by:
\begin{eqnarray}
    \rho_Q&=&-T_0^0=\frac{1}{2}\dot{\phi}^2+V\left(\phi\right),\label{64} \\
    p_Q&=&T_i^i=\frac{1}{2}\dot{\phi}^2-V\left(\phi\right).\label{65}
\end{eqnarray}
Moreover, the EoS parameter $\omega_Q$ for the quintessence scalar field is given by:
\begin{eqnarray}
\omega_Q=\frac{p_Q}{\rho_Q}=\frac{\dot{\phi}^2-2V\left(\phi\right)}{\dot{\phi}^2+2V\left(\phi\right)}.\label{66}
\end{eqnarray}
We now derive from Eq. (\ref{66}) that, when $\omega_Q < -1/3$, the universe accelerates if the condition $\dot{\phi}^2<V\left(\phi\right)$ is satisfied. Then,
the scalar potential needs to be shallow enough in order to have that the field evolves slowly along the potential. \\
The variation with respect to $\phi$ of the quintessence action given in Eq. (\ref{61}) yields to:
\begin{eqnarray}
    \ddot{\phi} + 3H\dot{\phi}+V_{,\phi} = 0,
\end{eqnarray}
where the two dots indicate the second derivative with respect to the cosmic time $t$, the single dot indicates the derivative with respect to the cosmic time $t$ and
$V_{,\phi} \equiv dV/d\phi$.\\
Many quintessence potentials have been already proposed. They have been usually classified into (i) freezing models and (ii) thawing models \cite{cald2005}. In the
freezing models, the field was rolling along the potential in the past, but the movement gradually slows down after the system enters the phase of cosmic acceleration.
In the thawing models,  the field (with mass $m_Q$) has been frozen by Hubble friction (i.e. the term $H\dot{\phi}$) until recently and then it begins to evolve once
$H$ drops below $m_Q$. The equation of state of DE is $\omega_Q\cong 1$ at early times, which is followed by the growth of $\omega_Q$.\\
Here we establish the correspondence between the interacting scenario and the quintessence DE model: equating Eq. (\ref{66}) with the EoS parameter (\ref{23}) (i.e.
$\omega_Q=\omega_{\Lambda}$) and Eq. (\ref{64}) with Eq. (\ref{5}) (i.e. $\rho_Q=\rho_{\Lambda}$), we obtain the following expressions for the the kinetic energy term
$\dot{\phi}^2$ and the quintessence potential energy $V\left(\phi \right)$:
\begin{eqnarray}
    \dot{\phi}^2&=&\left(1+\omega_{\Lambda} \right)\rho_{\Lambda}, \label{67}\\
    V\left( \phi \right) &=& \frac{1}{2}\left(1-\omega_{\Lambda} \right)\rho_{\Lambda}.\label{68}
\end{eqnarray}
Substituting the EoS parameter given in Eq. (\ref{23}) into Eqs. (\ref{67}) and (\ref{68}) we obtain:
\begin{eqnarray}
    \dot{\phi}^2&=&\rho_{\Lambda}\left(1-\frac{1}{3\left(3\gamma -\beta R^{\frac{\alpha}{2}-1}\right)} + \frac{\left(1+ \Omega _k  \right)}{3\Omega _{\Lambda}}\right),\label{69} \\
    V\left( \phi \right) &=&\frac{\rho_{\Lambda}}{2}  \left( 1+\frac{1}{3\left(3\gamma -\beta R^{\frac{\alpha}{2}-1}\right)} - \frac{\left(1+
\Omega _k  \right)}{3\Omega _{\Lambda}} \right).\label{70}
\end{eqnarray}
We can now obtain the evolutionary form of the quintessence scalar field integrating Eq. (\ref{69}) with respect to the scale facotr $a$ and using the
relation $\dot{\phi}=\phi' H$:
\begin{eqnarray}
\phi\left(a\right) - \phi \left(a_0\right)
=
\int_{a_0}^{a}\frac{da}{a}\sqrt{3M_p^2\Omega_{\Lambda}\left(1-\frac{1}{3(3\gamma
-\beta R^{\frac{\alpha}{2}-1})} + \frac{\left(1+ \Omega _k
\right)}{3\Omega _{\Lambda}}\right)},\label{71}
\end{eqnarray}
where $a_0$ is the present value of the scale factor.\\
In the limiting case for flat dark dominated universe, i.e. when $\beta=0$, $\Omega_{\Lambda}=1$ and $\Omega_k$=0, the scalar field and potential of quintessence
reduces, respectively, to:
\begin{eqnarray}
\phi(t)=\frac{6\gamma M_p}{\sqrt{3\gamma(12\gamma-1)}}\ln{(t)},\label{72}
\end{eqnarray}
\begin{eqnarray}
V(\phi)=\frac{6\gamma(6\gamma+1)}{(12\gamma-1)^2}M_p^2\exp{\left[\frac{-\sqrt{3\gamma(12\gamma-1)}}{3\gamma
M_p}\phi\right]}.\label{73}
\end{eqnarray}
The potential exists for all values of $\gamma >1/12$ (which correspond to the quitessence regime). The potential has also been obtained by power-law expansion of the scale factor.

\subsection{MODIFIED  CHAPLYGIN GAS (MCG)}
In this Section we want to obtain a correspondence between the Modified Chaplygin Gas (MCG) and the R-PLECHDE model.\\
One of the suggested candidates for DE is the Generalized Chaplygin Gas (GCG), which represents the generalization of the Chaplygin Gas \cite{mcg1}. GCG has the
favourable property of interpolating the evolution of the universe from the dust to the accelerated phase, hence it fits better the observational data \cite{mcg2}.
The GCG and its further generalization have been widely studied in literature \cite{mcg3}. \\
The GCG is defined as \cite{30}:
\begin{eqnarray}
    p_{\Lambda}=-\frac{D}{\rho_{\Lambda}^{\theta}}, \label{gcg1}
\end{eqnarray}
where $D$ and $\theta$ are two constants ($D$ is also positive defined). The Chaplygin gas is obtained in the limiting case $\theta = 1$. \\
The Modified Chaplygin Gas (MCG) represents a generalization of the GCG with the addition of a barotropic term. The MCG seems to be consistent
with the 5-year Wilkinson Microwave Anisotropy Probe (WMAP) data and henceforth support the unified model with DE and matter based on generalized Chaplygin gas.\\
The MCG is defined as \cite{33}:
\begin{equation}
p_{\Lambda}=A\rho_{\Lambda}-\frac{D}{\rho_{\Lambda}^\theta}, \label{mcg1}
\end{equation}
where $A$ and $D$ are two positive constants and $0 \leq \theta \leq 1$.\\
The density evolution of the MCG, calculated using the density conservation equation, is given by:
\begin{equation}
\rho_{\Lambda}=\left[\frac{D}{A+1}+\frac{B}{a^{3(\theta+1)(A+1)}}\right]^{\frac{1}{\theta+1}},\label{mcg2}
\end{equation}
where $B$ represents a constant of integration.\\
We now want to reconstruct the potential and dynamics of the scalar field $\Phi$ in the light of R-PLECHDE. For a homogeneous and time dependent scalar field
$\Phi$, energy density and pressure are defined, respectively, by:
\begin{eqnarray}
\rho_{\Lambda}&=&\frac{\sigma}{2}\dot\Phi^2+V(\Phi),\label{mcg3}\\
p_{\Lambda}&=&\frac{\sigma}{2}\dot\Phi^2-V(\Phi).\label{mcg4}
\end{eqnarray}
The case with $\sigma = -1$ corresponds to the phantom, instead the case with $\sigma = +1$ corresponds to the standard scalar field which represents the quintessence
field. Moreover, $V (\phi)$ represents the scalar potential of the field.\\
The EoS paramater $\omega_{\Lambda}$ of the MCG is given by:
\begin{equation}
\omega_{\Lambda}=\frac{p_{\Lambda}}{\rho_{\Lambda}}=\frac{\sigma \dot\Phi^2-2V(\Phi)}
{\sigma \dot\Phi^2+2V(\Phi)}.\label{mcg5}
\end{equation}
Using Eqs. (\ref{mcg3}), (\ref{mcg4}) and (\ref{mcg5}), we get the kinetic energy $\dot{\Phi}^2$ and the scalar potential $V\left(\Phi\right)$ terms, respectively, as:
\begin{eqnarray}
\dot{\Phi}^2&=&\frac{1}{\sigma}(1+\omega_{\Lambda})\rho_{\Lambda},\label{mcg6}\\
V\left(\Phi\right)&=&\frac{1}{2}(1-\omega_\Lambda)\rho_{\Lambda}.\label{mcg7}
\end{eqnarray}
We also know that the EoS paramater $\omega_{\Lambda}$ can be written as:
\begin{equation}
\omega_{\Lambda}=A-\frac{D}{\rho_{\Lambda}^{\theta +1}}.\label{mcg8}
\end{equation}
From Eq. (\ref{mcg2}), we can easily derive that:
\begin{eqnarray}
B=a^{3(\theta +1)(A+1)}\left(\rho _{\Lambda }^{\theta+1}-\frac{D}{A+1}\right). \label{mcgb}
\end{eqnarray}
Moreover, from Eq. (\ref{mcg8}), we obtain the following relation for $D$:
\begin{equation}
D=\rho_{\Lambda}^{\theta +1}\left(A-\omega_{\Lambda}\right). \label{mcg9}
\end{equation}
Substistuing in Eq. (\ref{mcgb}) the expression of $D$ given in Eq (\ref{mcg9}), we obtain that $B$ can be rewritten as:
\begin{eqnarray}
B=\left(a^{3\left(A+1\right)}\rho_{\Lambda}\right)^{1+\theta}   \left( \frac{1+\omega_{\Lambda}}{1+A} \right).\label{MMmcg}
\end{eqnarray}
Inserting the EoS parameter $\omega_{\Lambda}$ of the R-PLECHDE given in Eq. (\ref{23}) into Eq. (\ref{mcg9}), we derive:
\begin{eqnarray}
D =\left[\rho_{\Lambda}\right]^{\theta +1}\left [A +\frac{1}{3\left(3\gamma -\beta R^{\frac{\alpha} {2}-1}\right)}-\frac{1}{3}\left(\frac{1+\Omega_k}{\Omega_{\Lambda}}\right)
\right].\label{mcg11}
\end{eqnarray}
Instead, inserting in Eq. (\ref{MMmcg}) the EoS parameter $\omega_{\Lambda}$ of the R-PLECHDE given in Eq. (\ref{23}), we have:
\begin{eqnarray}
B =\frac{[a^{3(A+1)}\rho_{\Lambda}]^{\theta
+1}}{1+A}\left[1- \frac{1}{3\left(3\gamma -\beta R^{\frac{\alpha} {2}-1}\right)}+\frac{1}{3}\left(\frac{1+\Omega_k}{\Omega_{\Lambda}}
  \right)\right].\label{mcg13}
\end{eqnarray}
Using Eqs. (\ref{mcg6}), (\ref{mcg7}), (\ref{mcg11}) and (\ref{mcg13}), we obtain that the kinetic and potential terms for the R-PLECHDE model can be written as:
\begin{eqnarray}
\dot\Phi^2 &=&\frac{\rho_{\Lambda}}{\sigma}\left[1-\frac{1}{3\left(3\gamma -\beta R^{\frac{\alpha} {2}-1}\right)}+\frac{1}{3}\left(\frac{1+\Omega_k}{\Omega_{\Lambda}}
  \right)  \right],\label{mcg14-1}
\end{eqnarray}
\begin{eqnarray}
V(\Phi)&=&\frac{1}{2}\left[1 +\frac{1}{3\left(3\gamma -\beta R^{\frac{\alpha} {2}-1}\right)}-\frac{1}{3}\left(\frac{1+\Omega_k}{\Omega_{\Lambda}}\right)
\right] .\label{mcg15-1}
\end{eqnarray}
We can find the evolutionary form of the MCG scalar field integrating Eq. (\ref{mcg14-1}) with respect to the scale factor $a$:
\begin{eqnarray}
\Phi\left(a\right) - \Phi\left(a_0\right) &=& \nonumber \\   &=&\int_{a_0}^{a}\left\{\left[\frac{3M_p^2\Omega_{\Lambda}}{\sigma}
\left(1- \frac{1}{3\left(3\gamma -\beta R^{\frac{\alpha} {2}-1}\right)}+\frac{1}{3}\left(\frac{1+\Omega_k}{\Omega_{\Lambda}}\right)\right)\right]\right\}^{1/2} \frac{da}{a}, \label{mcg17}
\end{eqnarray}
where we used the relation $\dot{\Phi}=\Phi' H$ and $a_0$ is the present value of the scale factor.\\
In the limiting case for flat dark dominated universe, i.e. $\beta=0$, $\Omega_{\Lambda}=1$ and $\Omega_k$=0,
the scalar field $\phi\left(t \right)$ and the potential $V(\phi)$ of the MCG reduce, respectively, to:
\begin{eqnarray}
    \phi\left(t \right) = \frac{6\gamma M_p}{\sqrt{3\gamma \sigma \left( 12\gamma -1\right)}}\ln \left( t \right), \label{mcg22}
\end{eqnarray}
\begin{eqnarray}
V(\phi)=\frac{6\gamma(6\gamma+1)}{(12\gamma-1)^2}M_p^2\exp{\left[\frac{-\sqrt{3\gamma(12\gamma-1)}}{3\gamma
M_p}\phi\right]}.\label{mcg23}
\end{eqnarray}

\section{Conclusion}
In this paper, we studied the entropy-corrected version of the HDE model which
is in interaction with DM in the non-flat FRW universe. We considered as IR cut-off
the Ricci scalar $R$. The HDE model is an attempt to probe the nature of DE within
the framework of quantum gravity. We considered the power-law corrected term to the
energy density of HDE model. Using the expression of the modified energy
density, we obtained the EoS parameter $\omega_{\Lambda}$, the deceleration parameter $q$ and the evolution of energy
density parameter $\Omega_D'$ for the interacting R-PLECHDE model. We found that, for the appropriate model
parameters (even in the limiting case for flat dark dominated universe, i.e. $\beta=0$, $\Omega_{\Lambda}=1$ and $\Omega_k$=0),
the phantom divide may be crossed, i.e. $\omega_{\Lambda} < 1$, and the present acceleration expansion ($q<0$) is achieved where the quintessence is started.
Moreover, we established a correspondence between the interacting R-PLECHDE model and the Modified Chaplygin Gas (MCG) and the tachyon, K-essence, dilaton and
quintessence scalar fields in the hypothesis of non-flat FRW universe.\\
These correspondences are important to understand how various candidates of DE are
mutually related to each other. The limiting case of flat dark dominated universe without
entropy correction were studied in each scalar field and we see that the EoS parameter
is constant in this case and we calculate the scalar field and its potential which can be
obtained by idea of power-law expansion of scalar field.\\
In order to make a comparison between the R-PLECHDE and other works in PLECHDE-scalar
field model, we concentrate our attention in two papers recently written, one by Granda and Oliveros in 2009 \cite{granda} and one by Khodam-Mohammadi in 2011 \cite{khodam}.
Granda and Oliveros introduced an infrared cut-off which is function of the Hubble parameter $H$ and the derivative of the Hubble parameter
$\dot{H}$ with respect to the cosmic time, i.e. $L_{GO} = \left( \alpha H^2 + \beta\dot{H}  \right)^{-1/2}$, where $\alpha$ and $\beta$ are two constant. In the limiting case of $\alpha=2$ and $\beta=1$, $L_{GO}$ is equal to the Ricci scalar in the case the curvature parameter $k$ is equal to zero. Khodam-Mohammadi studied the power-law entropy corrected HDE model using as infrared cut-off $L_{GO}$. The results obtained in this work are in good agreement with those obtained in both works in the limiting case of flat
dark dominated universe.

\end{document}